\documentclass[superscriptaddress,notitlepage]{revtex4-1}
\usepackage[english]{babel}
\usepackage[utf8]{inputenc}
\usepackage{float}
\usepackage{color}
\usepackage{amsmath,amssymb,graphicx,textcomp}

\newcommand{\tmop}[1]{\ensuremath{\operatorname{#1}}}

\begin{document}

\title{Using infinite volume, continuum QED and lattice QCD for the hadronic light-by-light contribution
to the muon anomalous magnetic moment}

\newcommand{\RBRC}{
  RIKEN BNL Research Center,
  Brookhaven National Laboratory,
  Upton, New York 11973,
  USA}

\newcommand{\UCONN}{
  Physics Department,
  University of Connecticut,
  Storrs, Connecticut 06269-3046,
  USA}

\newcommand{\NAGOYA}{
  Department of Physics,
  Nagoya University,
  Nagoya 464-8602,
  Japan}

\newcommand{\NISHINA}{
  Nishina Center,
  RIKEN,
  Wako, Saitama 351-0198,
  Japan}

\newcommand{\BNL}{
  Physics Department,
  Brookhaven National Laboratory,
  Upton, New York 11973,
  USA}

\newcommand{\CU}{
  Physics Department,
  Columbia University,
  New York, New York 10027,
  USA}

\newcommand{\KEK}{
  KEK Theory Center,
  Institute of Particle and Nuclear Studies,
  High Energy Accelerator Research Organization (KEK),
  Tsukuba 305-0801,
  Japan
}

\author{Thomas Blum}
\affiliation{\UCONN}
\affiliation{\RBRC}

\author{Norman Christ}
\affiliation{\CU}

\author{Masashi Hayakawa}
\affiliation{\NAGOYA}

\author{Taku Izubuchi}
\affiliation{\BNL}
\affiliation{\RBRC}

\author{Luchang Jin}
\email{ljin.luchang@gmail.com}
\affiliation{\BNL}

\author{Chulwoo Jung}
\affiliation{\BNL}

\author{Christoph Lehner}
\affiliation{\BNL}

\begin{abstract}
  In Ref {\cite{Blum:2016lnc}\nocite{Blum:2015gfa}}, the connected and leading
  disconnected hadronic light-by-light contributions to the muon anomalous magnetic moment ($g - 2$) have been computed
  using lattice QCD ensembles corresponding to physical pion mass generated by the RBC/UKQCD
  collaboration. However, the calculation is expected to suffer from
  a significant finite volume error that scales like $1 / L^2$ where $L$
  is the spatial size of the lattice.
  In this paper, we demonstrate that this problem is cured by treating the muon and photons in infinite volume, continuum QED,
  resulting in a weighting function that is pre-computed and saved
  with affordable cost and sufficient accuracy.
  {We present numerical results for the case when
  the quark loop is replaced by a muon loop, finding
  the expected exponential approach to the infinite
  volume limit and consistency with the known analytic
  result.\nocite{Asmussen:2016lse,Green:2015sra}
  }
  We have implemented an improved weighting function which reduces both discretization
  and finite volume effects arising from the hadronic part of the amplitude.
\end{abstract}

\maketitle

\section{Introduction}

Precision measurements of lepton magnetic dipole moments provide a
powerful tool to test the standard model (SM) of particle physics at high
precision.  The magnetic dipole moment $\vec\mu$ originating from the
lepton's spin $\vec{s}$ is commonly expressed as
\begin{align}
  \vec{\mu} = g
\left( \frac{e}{2m} \right) \vec{s} \,,
\end{align}
where $e$ is the lepton's electromagnetic charge and $m$ is its mass.
The anomalous magnetic moment, or anomaly, $a=(g-2)/2$ expresses the
deviation from Dirac's relativistic quantum-mechanical prediction
$g=2$.  It is generated by small radiative corrections which by a careful
comparison between its experimental measurement to its theory
prediction may reveal physics beyond the standard model.
Experimental measurements have determined these anomalous moments at
very high precision.  The electron anomaly, $ a_e =
0.00115965218073(28)$ \cite{Hanneke:2008tm}, currently yields the
most precise value of the fine structure constant $ \alpha=
1/137.035999157(33)$ \cite{Aoyama:2014sxa}.
In general, contributions from a new physics scale $\Lambda_{\rm NP}$
to the anomalous magnetic moment of a lepton $\ell=e,\mu,\tau$ are
suppressed by $m^2_\ell / \Lambda^2_{\rm NP}$.  One therefore expects
the muon to be five orders of magnitude more sensitive to such
contributions than the electron which outweighs a loss in experimental
precision.  With the $\tau$ being experimentally inaccessible, $a_\mu$
is the most promising channel to reveal physics beyond the standard
model.

Interestingly, current experimental and theoretical determinations of $a_\mu$
differ at the $3.1$--$3.5$ standard deviation level,
\begin{align}
  a_\mu^{\rm EXP} - a_\mu^{\rm SM} = \,\,& (27.6 \pm 8.0) \times 10 ^ {-10}~\text{\cite{Davier:2010nc}} \,, \notag\\
  &(25.0 \pm 8.0) \times 10 ^ {-10}~\text{\cite{Hagiwara:2011af}} \,,
\end{align}
depending on which value for the hadronic vacuum polarization contribution
is used (see Tab.~\ref{tab:mugmtwo}).

\begin{table}[tb]
  \centering
  \begin{tabular}{lrr}\hline\hline
    Contribution & Value $\times 10^{10}$ & Uncertainty $\times 10^{10}$\\\hline
    QED & 11 658 471.895 & 0.008 \\
    Electroweak Corrections & 15.4 & 0.1 \\
    HVP (LO) \cite{Davier:2010nc} & 692.3  & 4.2 \\
    HVP (LO) \cite{Hagiwara:2011af} & 694.9  & 4.3 \\
    HVP (NLO) & -9.84 & 0.06 \\
    HVP (NNLO) & 1.24 & 0.01 \\
    HLbL & 10.5 & 2.6 \\
    \hline
    Total SM prediction \cite{Davier:2010nc} & 11 659 181.5 & 4.9 \\
    Total SM prediction \cite{Hagiwara:2011af} & 11 659 184.1 & 5.0 \\    \hline
    BNL E821 result & 11 659 209.1 & 6.3 \\
    Fermilab E989 target & & $\approx$  1.6\\\hline\hline
  \end{tabular}
  \caption{Individual contributions to the current standard model calculation of $a_\mu$ \cite{PDG2013,Kurz:2014wya}.
    The BNL E821 experimental result \cite{Bennett:2006fi} and Fermilab E989 target precision \cite{Grange:2015fou}
    are given for comparison.}
  \label{tab:mugmtwo}
\end{table}

In this tension the theory and experimental uncertainties are approximately balanced,
with the theory uncertainty dominated by the hadronic
vacuum polarization and hadronic light-by-light (HLbL) contributions.
With future experiments at Fermilab (E989) \cite{Carey:2009zzb}
and J-PARC (E34) \cite{Aoki:2009xxx} aiming for a four-fold decrease
in experimental uncertainty, a careful first-principles determination of these
hadronic contributions and a similar reduction in uncertainty is desirable.

In this work we present an improved method to compute the HLbL
contribution from first principles in lattice quantum chromodynamics
(QCD).  We build on the optimized sampling strategy of the HLbL
diagrams, which we have introduced in Ref.~\cite{Blum:2015gfa} and
which has reduced the statistical uncertainties, at the same cost, by more
than an order of magnitude compared to the pioneering work of
Ref.~\cite{Blum:2014oka}.  In a recent publication
\cite{Blum:2016lnc}, we have presented a first-principles 2+1 flavor lattice QCD
calculation of the connected and leading disconnected contributions to the muon anomaly at
physical quark and muon masses,
\begin{align}
  a_\mu^{\rm HLbL} &= 5.35(1.35) \times 10^{-10} \,,
\end{align}
where the statistical uncertainty is given.  This result is affected
by potentially large systematic errors due to the non-zero lattice spacing and the
finite lattice volume used in our calculation.  We are in the process
of repeating our calculation on a second lattice spacing to address the
former systematic.  The latter is addressed in this work.

So far all lattice QCD calculations of the HLbL contribution to the muon $g-2$ have treated the photons and muon in the same finite hypercubic lattice where the quarks and gluons live.
The results are expected to suffer from sizable finite volume corrections which scale as some power of the system size rather than the exponential scaling observed for typical lattice QCD calculations since the photons are restricted to a finite box.
Inspired by earlier work on the hadronic vacuum polarization \cite{Blum:2002ii},
we remove power-law finite volume errors by computing the muon and photon
components of our diagrams in infinite volume and subsequently combine the resulting
weight function with a QCD four-point function obtained in our lattice simulation.
{
The Mainz group announced a similar approach \cite{Asmussen:2016lse,Green:2015sra},
which, to a large extent, motivated this work.
}

In the following we describe our method in detail and verify it in
the leptonic case, where we replace the quark by a lepton loop.  This
replacement is trivial from the perspective of our lattice calculation
and the same setup with free propagators replaced by propagators on a
non-trivial QCD background allows us to perform the calculation in the
desired QCD case.

\begin{figure}[H]
  \begin{center}
    \resizebox{0.4\columnwidth}{!}{\includegraphics{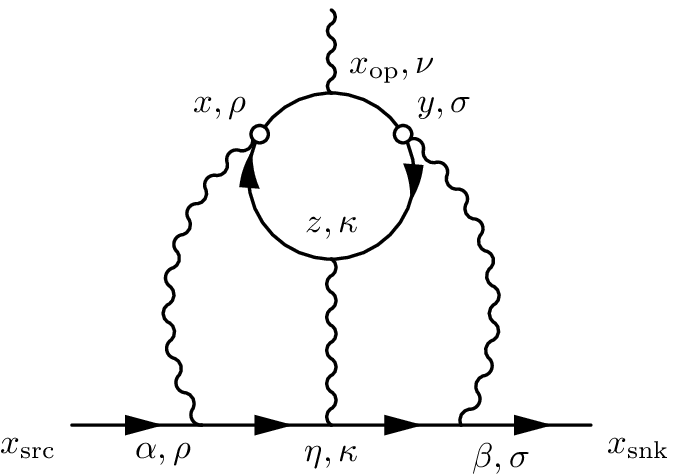}}
    \
    \resizebox{0.4\columnwidth}{!}{\includegraphics{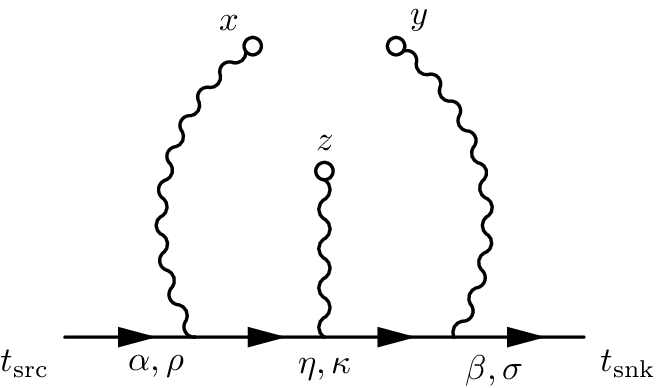}}
  \end{center}
  \caption{\label{fig:clbl}The connected light-by-light diagram. There are five
  other diagrams like the one on the left that correspond to distinct ways of connecting the photons to the muon line (or equivalently, the quark loop).}
\end{figure}

In Ref {\cite{Blum:2015gfa}}, we introduced a formula to obtain the connected
hadronic light-by-light contribution to the anomalous magnetic moment {given by the electromagnetic Pauli form factor evaluated at zero momentum transfer}, $F_2 (q^2 =
0)$, from a lattice calculation:
\begin{eqnarray}
  \frac{F_2^{\tmop{cHLbL}} (q^2 = 0)}{m}  \frac{(\sigma_{s', s})_i}{2} & = &
  \frac{1}{VT} \sum_{x, y, z} \sum_{x_{\text{op}}}  \frac{1}{2} \epsilon_{i,
  j, k} \left( x_{\text{op}} - x_{\tmop{ref}} \right)_j \cdot i \bar{u}_{s'}
  (\vec{0}) \mathcal{F}^C_k \left( x, y, z, x_{\text{op}} \right) u_s
  (\vec{0}),  \label{eq:f2-lbl-moment}
\end{eqnarray}
where $(\sigma_{s', s})_i = \bar{u}_{s'} (\vec{0}) \Sigma_i u_s (\vec{0})$ are
the conventional Pauli matrices.
The coordinates $x_\text{op}, ~x,~y,~z$ are the locations of the electromagnetic currents on the quark loop, the former corresponding to the external photon and the latter to the virtual photons connecting the quark loop to the muon (see Fig.~\ref{fig:clbl}).
The point $x_{\tmop{ref}}$ can be chosen arbitrarily
and may even depend on $x$, $y$, and $z$. In Ref {\cite{Blum:2015gfa}}, we set
$x_{\tmop{ref}} = (x + y) / 2$ and further manipulated the above formula to
take advantage of the symmetry between $x$, $y$, $z$ to reduce the
statistical noise inherent in our monte carlo integration:
\begin{eqnarray}
  \frac{F_2^{\tmop{cHLbL}} (q^2 = 0)}{m}  \frac{(\sigma_{s', s})_i}{2}
  & = & \sum_{r, \tilde{z}} \mathfrak{Z} \left( \frac{r}{2}, - \frac{r}{2},
  \tilde{z} \right) \sum_{\tilde{x}_{\text{op}}}  \frac{1}{2} \epsilon_{i, j,
  k} \left( \tilde{x}_{\text{op}} \right)_j \cdot i \bar{u}_{s'} (\vec{0})
  \mathcal{F}^C_k \left( \frac{r}{2}, - \frac{r}{2}, \tilde{z},
  \tilde{x}_{\text{op}} \right) u_s (\vec{0}).
  \label{eq:f2-lbl-moment-short-z}
\end{eqnarray}
The integration variables are related to the coordinates in Fig.
\ref{fig:clbl} by the following equations: $r = x - y$, $\tilde{z} = z - (x +
y) / 2$, $\tilde{x}_{\text{op}} = x_{\text{op}} - (x + y) / 2$. The function
``$\mathfrak{Z}$'' is defined by
\begin{eqnarray}
  \mathfrak{Z} (x, y, z) & = & \left\{ \begin{array}{ll}
    3 & \text{if } | x - y | < | x - z | \text{ and } | x - y | < | y - z |\\
    3 / 2 & \text{if } | x - y | = | x - z | < | y - z | \text{ or } | x - y |
    = | y - z | < | x - z |\\
    1 & \text{if } | x - y | = | x - z | = | y - z |\\
    0 & \text{otherwise}
  \end{array} \right. .  \label{eq:3-mult}
\end{eqnarray}
We compute the summation over $r$ in Eq.~(\ref{eq:f2-lbl-moment-short-z}) by
stochastically sampling $x$ and $y$ point pairs, while the sums over
$\tilde{x}_{\text{op}}$ and $\tilde{z}$ are performed completely over the entire lattice.
The amplitude $\mathcal{F}^C_{\nu} \left( x, y, z, x_{\text{op}} \right)$ is
given by:
\begin{eqnarray}
  \mathcal{F}^C_{\nu} \left( x, y, z, x_{\text{op}} \right) & = & (- i e)^6
  \mathcal{G}_{\rho, \sigma, \kappa} (x, y, z) \mathcal{H}^C_{\rho, \sigma,
  \kappa, \nu} (x, y, z, x_{\tmop{op}}),  \label{eq:lbl-amp}
\end{eqnarray}
where $i^4 \mathcal{H}^C_{\rho, \sigma, \kappa, \nu} (x, y, z, x_{\tmop{op}})$
represents the four-point hadronic correlation function, and $i^3 \mathcal{G}_{\rho,
\sigma, \kappa} (x, y, z)$ is the QED weighting function.
For the connected diagram, $i^4 \mathcal{H}^C_{\rho,
\sigma, \kappa, \nu} (x, y, z, x_{\tmop{op}})$ is given by the following two
equations:
\begin{eqnarray}
  i^4 \mathcal{H}^C_{\rho, \sigma, \kappa, \nu} (x, y, z, x_{\tmop{op}}) & = &
  \frac{1}{6} \mathcal{H}_{\rho, \sigma, \kappa, \nu} (x, y, z, x_{\tmop{op}})
  + \frac{1}{6} \mathcal{H}_{\sigma, \kappa, \rho, \nu} (y, z, x,
  x_{\tmop{op}}) + \frac{1}{6} \mathcal{H}_{\kappa, \rho, \sigma, \nu} (z, x,
  y, x_{\tmop{op}}) \\
  &  & + \frac{1}{6} \mathcal{H}_{\kappa, \sigma, \rho, \nu} (z, y, x,
  x_{\tmop{op}}) + \frac{1}{6} \mathcal{H}_{\rho, \kappa, \sigma, \nu} (x, z,
  y, x_{\tmop{op}}) + \frac{1}{6} \mathcal{H}_{\sigma, \rho, \kappa, \nu} (y,
  x, z, x_{\tmop{op}}), \nonumber
\end{eqnarray}
\begin{eqnarray}
  \mathcal{H}_{\rho, \sigma, \kappa, \nu} (x, y, z, x_{\tmop{op}}) & = &
  \sum_{q = u, d, s} (e_q / e)^4 \left\langle - \tmop{tr} \left[ i
  \gamma_{\rho} S_q (x, z) i \gamma_{\kappa} S_q (z, y) i \gamma_{\sigma} S_q
  \left( y, x_{\text{op}} \right) i \gamma_{\nu} S_q \left( x_{\text{op}}, x
  \right) \right] \right\rangle_{\text{QCD}} ,  \label{eq:hadronic-loop}
\end{eqnarray}
where $e_u / e = 2 / 3$, and $e_d / e = e_s / e = - 1 / 3$. The QED weighting
function, $i^3 \mathcal{G}_{\rho, \sigma, \kappa} (x, y, z)$, is a symmetrized
version of $\mathfrak{G}_{\sigma, \kappa, \rho} (y, z, x)$, which is
represented by the right diagram of Fig. \ref{fig:clbl}:
\begin{eqnarray}
  i^3 \mathcal{G}_{\rho, \sigma, \kappa} (x, y, z) & = & \mathfrak{G}_{\rho,
  \sigma, \kappa} (x, y, z) +\mathfrak{G}_{\sigma, \kappa, \rho} (y, z, x)
  +\mathfrak{G}_{\kappa, \rho, \sigma} (z, x, y)  \label{eq:muon-line-sym}\\
  &  & +\mathfrak{G}_{\kappa, \sigma, \rho} (z, y, x) +\mathfrak{G}_{\rho,
  \kappa, \sigma} (x, z, y) +\mathfrak{G}_{\sigma, \rho, \kappa} (y, x, z),
  \nonumber
\end{eqnarray}
\begin{eqnarray}
  \mathfrak{G}_{\sigma, \kappa, \rho} (y, z, x) & = & \lim_{t_{\text{src}}
  \rightarrow - \infty, t_{\text{snk}} \rightarrow \infty} e^{m_{{\mu}}
  \left( t_{\text{snk}} - t_{\text{src}} \right)} \int_{\alpha, \beta, \eta} G
  (x, \alpha) G (y, \beta) G (z, \eta)  \label{eq:muon-line}\\
  &  & \times \int_{\vec{x}_{\text{snk}}, \vec{x}_{\text{src}}} S_{{\mu}}
  \left( x_{\text{snk}}, \beta \right) i \gamma_{\sigma} S_{{\mu}} (\beta,
  \eta) i \gamma_{\kappa} S_{{\mu}} (\eta, \alpha) i \gamma_{\rho}
  S_{{\mu}} \left( \alpha, x_{\text{src}} \right), \nonumber
\end{eqnarray}
where $S_{{\mu}}$ and $G$ are free muon and photon propagators
respectively.

In the past, we evaluated the QED weighting function, $i^3 \mathcal{G}_{\rho,
\sigma, \kappa} (x, y, z)$, on a finite size lattice, which resulted in $1 /
L^2$ finite volume errors, where $L$ is the size of the lattice that is used
to evaluate $i^3 \mathcal{G}_{\rho, \sigma, \kappa} (x, y, z)$
{\cite{Blum:2015gfa}}. This lattice was referred to as the QED box. Although
one can make the QED box much larger than the QCD box {\cite{Jin:2015bty}}, it
is far better to compute the QED weighting function in infinite
volume (and in the continuum) directly as proposed in Ref. {\cite{Asmussen:2016lse,Green:2015sra}}.

{
It may be useful to recall the finite-volume effects expected in the calculation of the hadronic light-by-light scattering
contributions we are studying.
The mass gap of QCD has two implications for a hadronic correlation function such as
$\mathcal{H}^C_{\rho, \sigma, \kappa, \nu} (x, y, z, x_{\tmop{op}})$: %
(1) The correlation function will decrease exponentially as the space-time distances between its
arguments grow; (2) For fixed locations of its arguments, the finite-volume errors in such a correlation function will fall
exponentially in the linear size of the volume in which it is computed. Therefore, if we evaluate the QED weighting
function in infinite volume, which is the focus of this paper, and keep the positions $x; y; z; x_\text{op}$ fixed, all finite-volume
errors will be exponentially suppressed as the linear lattice size grows. Since the QED weighting does not grow
exponentially when the separations between $x, y, z$ increase, the summation in Eq. (\ref{eq:f2-lbl-moment-short-z})
converges exponentially implying
that all finite-volume errors in the result for the muon anomaly are exponentially suppressed. For the same reason, one
concludes that the finite-volume errors for the lattice calculation of the hadronic vacuum polarization (HVP) contribution to
the muon $g-2$ [16] decrease exponentially as the lattice volume is increased. This conclusion will remain true when the
QED corrections are included, if they are treated by a method similar to that used here.
One should keep in mind that the use of an infinite-volume photon propagator in other contexts may not achieve the
same reduction of finite-volume errors in the HLBL case studied here.
}

In this work, we demonstrate our method to compute the QED weighting function
in infinite volume which differs significantly from the one proposed in Ref.
{\cite{Asmussen:2016lse}}. The paper is organized as follows:
In Sec. \ref{sec:formulation},
we perform some analytic calculations and reduce the 12
dimensional integration in $\mathfrak{G}_{\sigma, \kappa, \rho} (y, z, x)$ to a
four dimensional integration, which we then integrate numerically with {the CUBA library}
cubature rules {\cite{Hahn:2004fe}}.
{We also} introduce a subtraction for
$\mathfrak{G}_{\sigma, \kappa, \rho} (y, z, x)$ which
does not alter the final result for $F_2$ in the infinite volume and
continuum limits of the QCD part.
In Sec. \ref{sec:results}, we show
results of a pure QED light-by-light calculation carried out in a fashion similar to
that of Ref. {\cite{Blum:2015gfa}}, but using the new infinite volume QED
weighting function and compare the two.
{In addition, we}
demonstrate that the new subtracted QED
weighting function reduces the
remaining exponentially suppressed finite volume and $\mathcal{O}
(a^2)$ discretization errors for $F_2$.

\section{Formulation}\label{sec:formulation}

Here we show how $\mathfrak{G}_{\sigma,
\kappa, \rho} (y, z, x)$ is evaluated using Eq.~(\ref{eq:muon-line}) in infinite volume. As
usual, we work in Euclidean space time, and the free muon and photon propagator
take the form
\begin{eqnarray}
  S_{{\mu}} (x, y) & = & \int \frac{d^4 p}{(2 \pi)^4}  \frac{1}{i \not{p}
  + m} e^{ip \cdot (x - y)} = \left( -\not\hskip -1.5pt{\partial}_x + m \right) \int
  \frac{d^4 p}{(2 \pi)^4}  \frac{1}{p^2 + m^2} e^{ip \cdot (x - y)}, \\
  G (x, y) & = & \int \frac{d^4 p}{(2 \pi)^4}  \frac{1}{p^2} e^{ip \cdot (x -
  y)} = \frac{1}{4 \pi^2}  \frac{1}{(x - y)^2} .
\end{eqnarray}
The wall- source and sink muon propagators that create and annihilate muons at rest appropriate for our kinematic setup can be evaluated as
\begin{eqnarray}
  \lim_{t_{\text{snk}} \rightarrow \infty} e^{m_{{\mu}} t_{\text{snk}}}
  \int_{\vec{x}_{\text{snk}}} S_{{\mu}} \left( x_{\text{snk}}, \beta
  \right) & = & \frac{\gamma_0 + 1}{2} e^{m_{{\mu}} \beta_t} ,
  \label{eq:muon-line-snk}\\
  \lim_{t_{\text{src}} \rightarrow - \infty} e^{- m_{{\mu}}
  t_{\text{src}}} \int_{\vec{x}_{\text{src}}} S_{{\mu}} \left( \alpha,
  x_{\text{src}} \right) & = & \frac{\gamma_0 + 1}{2} e^{- m_{{\mu}}
  \alpha_t}.  \label{eq:muon-line-src}
\end{eqnarray}
The matrix $\frac{\gamma_0 + 1}{2}$ is a projection operator, so
\begin{eqnarray}
  \mathfrak{G}_{\sigma, \kappa, \rho} (y, z, x) & = & \frac{\gamma_0 + 1}{2}
  \mathfrak{G}_{\sigma, \kappa, \rho} (y, z, x) \frac{\gamma_0 + 1}{2} .
  \label{eq:muon-line-proj}
\end{eqnarray}
Since $m_{{\mu}}$ is the only relevant scale in this function, without
loss in generality, we set $m_{{\mu}} = 1$. Starting with Eqs.~
(\ref{eq:muon-line-snk})~and~(\ref{eq:muon-line-src}), we find
\begin{eqnarray}
  \lim_{t_{\text{snk}} \rightarrow \infty} e^{m_{{\mu}} \left(t_{\text{snk}} - \eta_t\right)}
  \int_{\beta} G (y, \beta) \int_{\vec{x}_{\text{snk}}} S_{{\mu}} \left(
  x_{\text{snk}}, \beta \right) i \gamma_{\sigma} S_{{\mu}} (\beta, \eta)
  & = & \frac{\gamma_0 + 1}{2} i \gamma_{\sigma} \left( - \not{\hskip -2pt \partial_y} +
  \gamma_0 + 1 \right) f (\eta - y),
  \label{eq:muon-line-snk-part}\\
  \lim_{t_{\text{src}} \rightarrow - \infty} e^{- m_{{\mu}}
  \left(\eta_t - t_{\text{src}}\right)}
  \int_{\alpha} G (x, \alpha) \int_{\vec{x}_{\text{src}}}
  S_{{\mu}} (\eta, \alpha) i \gamma_{\rho} S_{{\mu}} \left( \alpha,
  x_{\text{src}} \right) & = & \left( \not{\hskip -2pt \partial_x} + \gamma_0 + 1 \right)
  i \gamma_{\rho}  \frac{\gamma_0 + 1}{2} f (x - \eta) ,
  \label{eq:muon-line-src-part}
\end{eqnarray}
where
\begin{eqnarray}
  f (x) = f (| x |, x_t / | x |) & = & \frac{1}{8 \pi^2} \int_0^1 dye^{- yx_t}
  K_0 (y | x |) ,
\end{eqnarray}
and $K_0(x)$ is a modified Bessel function of the second kind of order 0.
Next, substitute Eqs. (\ref{eq:muon-line-snk-part}) and (\ref{eq:muon-line-src-part}) into Eq.
(\ref{eq:muon-line}) to obtain
\begin{eqnarray}
  \mathfrak{G}_{\sigma, \kappa, \rho} (y, z, x) & = & \frac{\gamma_0 + 1}{2} i
  \gamma_{\sigma} \left( - \not{\hskip -2pt \partial_y} + \gamma_0 + 1 \right) i
  \gamma_{\kappa} \left( \not{\hskip -2pt \partial_x} + \gamma_0 + 1 \right) i
  \gamma_{\rho}  \frac{\gamma_0 + 1}{2}  \nonumber\\
  &  & \times \frac{1}{4 \pi^2} \int d^4 \eta \frac{1}{(\eta - z)^2} f (\eta
  - y) f (x - \eta) . \label{eq:muon-line-orig}
\end{eqnarray}
Before continuing to evaluate this function, let us prove some of its useful
properties. It should be noted that based on the definition, Eq.
(\ref{eq:muon-line}), the function $\mathfrak{G}_{\sigma, \kappa, \rho} (y, z,
x)$ has a logarithmic infrared divergence. This might raise concern about
whether it is correct to evaluate only the QED part of the light-by-light amplitude in infinite volume.
In fact we show below that the infrared divergence
can be avoided by using a new definition of the weighting function.

Recall that $\gamma_{{\mu}}$ and $S_{{\mu}} (x, y)$ are Hermitian Dirac
matrices which satisfy $\Sigma_2 \gamma_{{\mu}} \Sigma_2 =
(\gamma_{{\mu}})^T$ and $\Sigma_2 S_{{\mu}} (x, y) \Sigma_2 =
[S_{{\mu}} (x, y)]^T$. The free propagator is also translationally
invariant, so $S_{{\mu}} (x, y) = S_{{\mu}} (- y, - x)$. As a result,
one can show that
\begin{eqnarray}
  \Sigma_2 \mathfrak{G}_{\sigma, \kappa, \rho} (y, z, x) \Sigma_2 & = &
  [\mathfrak{G}_{\rho, \kappa, \sigma} (- x, - z, - y)]^T, \\
  {}[\mathfrak{G}_{\sigma, \kappa, \rho} (y, z, x)]^{\dagger} & = &
  -\mathfrak{G}_{\rho, \kappa, \sigma} (- x, - z, - y) .
  \label{eq:muon-line-hermitian}
\end{eqnarray}
It immediately follows that
\begin{eqnarray}
  \Sigma_2 \mathfrak{G}_{\sigma, \kappa, \rho} (y, z, x) \Sigma_2 & = & -
  [\mathfrak{G}_{\sigma, \kappa, \rho} (y, z, x)]^{\ast} .
\end{eqnarray}
Combining this result with Eq.~(\ref{eq:muon-line-proj}), we can parameterize
$\mathfrak{G}_{\sigma, \kappa, \rho} (y, z, x)$ as
\begin{eqnarray}
  \mathfrak{G}_{\sigma, \kappa, \rho} (y, z, x) & = & \frac{1 + \gamma_0}{2}
  [(a_{\sigma, \kappa, \rho} (y, z, x))_k \Sigma_k + ib_{\sigma, \kappa, \rho}
  (y, z, x)]  \frac{1 + \gamma_0}{2},  \label{eq:muon-line-param}
\end{eqnarray}
where $(a_{\sigma, \kappa, \rho} (y, z, x))_k$ and $b_{\sigma, \kappa, \rho}
(y, z, x)$ are real functions. Although the function $\mathfrak{G}_{\sigma,
\kappa, \rho} (y, z, x)$ is not Hermitian, the non-Hermitian part has no
projection to the magnetic moment. So, for the purpose of obtaining $F_2$, we
only need to evaluate its Hermitian component. Because we need to symmetrize
the arguments of the function in Eq.~(\ref{eq:muon-line-sym}), we can freely
permute the arguments of $\mathfrak{G}_{\sigma, \kappa, \rho} (y, z, x)$
without changing $F_2$. This allows us to define a new version of the function:
\begin{eqnarray}
  \mathfrak{G}^{(1)}_{\sigma, \kappa, \rho} (y, z, x) & = & \frac{1}{2}
  \mathfrak{G}_{\sigma, \kappa, \rho} (y, z, x) + \frac{1}{2}
  [\mathfrak{G}_{\rho, \kappa, \sigma} (x, z, y)]^{\dagger} \nonumber\\
  & = & \frac{1}{2} \mathfrak{G}_{\sigma, \kappa, \rho} (y, z, x) -
  \frac{1}{2} \mathfrak{G}_{\sigma, \kappa, \rho} (- y, - z, - x) .
  \label{eq:muon-line-v1}
\end{eqnarray}
As a special case, when all three coordinates are the same, we immediately
have
\begin{eqnarray}
  \mathfrak{G}^{(1)}_{\sigma, \kappa, \rho} (z, z, z) & = & 0,
  \label{eq:muon-line-zzz}
\end{eqnarray}
since $\mathfrak{G}_{\sigma, \kappa, \rho} (y, z, x)$ only depends on relative coordinates, or distance between its arguments.
Because the divergence of the function $\mathfrak{G}_{\sigma, \kappa, \rho}
(y, z, x)$ is infrared and logarithmic, it is independent of
the coordinates, $x$, $y$, $z$. One simple consequence of this behavior and Eq.~(\ref{eq:muon-line-zzz}) is that the new
version $\mathfrak{G}^{(1)}_{\sigma, \kappa, \rho} (y, z, x)$ is
infrared finite. Recall that the new version $\mathfrak{G}^{(1)}_{\sigma,
\kappa, \rho} (y, z, x)$ is the same as the original $\mathfrak{G}_{\sigma,
\kappa, \rho} (y, z, x)$ after substituting into Eq.~(\ref{eq:muon-line-sym}) and
projecting onto the magnetic moment. While the non-Hermitian part of the
original QED weighting function has a logarithmic infrared divergence, it does not contribute to the magnetic moment.

With Eqs. (\ref{eq:muon-line-orig}) and (\ref{eq:muon-line-v1}), we obtain an
infrared finite integration formula for $\mathfrak{G}^{(1)}_{\sigma, \kappa,
\rho} (y, z, x)$:
\begin{eqnarray}
  \mathfrak{G}^{(1)}_{\sigma, \kappa, \rho} (y, z, x) & = & \frac{\gamma_0 +
  1}{2} i \gamma_{\sigma} \left( \not{\hskip -2pt \partial_{\zeta}} + \gamma_0 + 1 \right)
  i \gamma_{\kappa} \left( \not{\hskip -2pt \partial_{\xi}} + \gamma_0 + 1 \right) i
  \gamma_{\rho}  \frac{\gamma_0 + 1}{2}   \label{eq:muon-line-v1-int}\\
  &  & \times \frac{1}{4 \pi^2} \int d^4 \eta \frac{1}{(\eta - z)^2}  \left.
  \frac{f (\eta - y + \zeta) f (x - \eta + \xi) - f (y - \eta + \zeta) f (\eta
  - x + \xi)}{2} \right|_{\xi = \zeta = 0}. \nonumber
\end{eqnarray}
This four dimensional integration is performed with the CUBA library's Cuhre
routine {\cite{Hahn:2004fe}}, which makes use of cubature rules and evaluates
the integration in a deterministic way. Since performing the numerical
integration is costly and the lattice calculation needs values
of this function for many different values of its arguments, we pre-compute $i^3
\mathcal{G}_{\rho, \sigma, \kappa} (x, y, z)$ for a range of points and
then approximate this function by interpolating the computed values,
which is similar to the strategy used in Ref. {\cite{Asmussen:2016lse}}.
The arguments of the function $i^3 \mathcal{G}_{\rho, \sigma, \kappa} (x, y, z)$
have 12 degrees of freedom. With the help of translation and
spatial rotational symmetries, the relevant number of degrees of freedom is
reduced to five. These five parameters are chosen to be
\begin{eqnarray}
  p_0 = (d / 6)^{1 / 2} & , \quad & d = | y - z |, \\
  p_1 = \alpha^{1 / 2} & , \quad & \alpha = | x - z | / d, \\
  p_2 = \theta / \pi & , \quad & \theta = \angle_{y - z, \hat{t}}, \\
  p_3 = \varphi / \pi & , \quad & \varphi = \angle_{x - z, \hat{t}}, \\
  p_4 = \eta / \pi & , \quad & \eta = \angle_{\overrightarrow{x - z},
  \overrightarrow{y - z}} .
\end{eqnarray}
Because $i^3 \mathcal{G}_{\rho, \sigma, \kappa} (x, y, z)$ is
symmetric with respect to permutation of its arguments, without loss of
generality, for the purpose of interpolation, we require $| y - z |
\geqslant | x - y | \geqslant | x - z |$
(this is unrelated to the restriction used for sampling the $x-y$ point pairs on the quark loop).
We also limit the length $d$ to
be less than $6$ (or roughly 11 fm) which should be large enough for the purpose of
computing the hadronic light-by-light diagrams on our lattices. We employ a straight forward
generalization of bilinear interpolation for the five dimensional
interpolation
({the interpolated function is linear with respect to any of its arguments
within the small region between the known data points}),
and the grid has uniform spacing in all directions with $0 \leqslant p_i \leqslant 1$.
We have computed {interpolation grids} with sizes
$6^5$, $8^5$, $10^5$, $12^5$, $14^5$, and $16^5$. In contrast to Ref.
{\cite{Asmussen:2016lse}}, we do not average over the muon propagation
direction, so our time direction is special. Thus we have a five- instead of a three-dimensional grid.
The two additional dimensions make the interpolation harder,
but as we shall see, the interpolation error remains under very good control.

{Although we introduced $\mathfrak{G}^{(1)}_{\sigma, \kappa, \rho} (y,
z, x)$ in addition to $\mathfrak{G}_{\sigma, \kappa, \rho} (y, z, x)$,
their differences will vanish immediately}
after projecting to the magnetic moment contribution and
substituting into Eq.~(\ref{eq:muon-line-sym}). However, due to
the conservation of electric-current in the hadronic four-point correlation function,
we enjoy even more freedom in choosing $\mathfrak{G}_{\sigma, \kappa, \rho} (y, z,
x)$. We introduce yet another version,
\begin{eqnarray}
  \mathfrak{G}^{(2)}_{\sigma, \kappa, \rho} (y, z, x) & = &
  \mathfrak{G}^{(1)}_{\sigma, \kappa, \rho} (y, z, x)
  -\mathfrak{G}^{(1)}_{\sigma, \kappa, \rho} (z, z, x)
  -\mathfrak{G}^{(1)}_{\sigma, \kappa, \rho} (y, z, z).
  \label{eq:muon-line-v2}
\end{eqnarray}
With this definition, $\mathfrak{G}^{(2)}_{\sigma, \kappa, \rho} (y, z, x)$
has the property that
\begin{eqnarray}
  \mathfrak{G}^{(2)}_{\sigma, \kappa, \rho} (z, z, x)
  =\mathfrak{G}^{(2)}_{\sigma, \kappa, \rho} (y, z, z) & = & 0.
  \label{eq:muon-line-v2-prop}
\end{eqnarray}
To demonstrate that these additional two terms in Eq.~(\ref{eq:muon-line-v2}) do not
contribute to the final result, recall the current conservation law for
$\mathcal{H}^C_{\rho, \sigma, \kappa, \nu} (x, y, z, x_{\tmop{op}})$:
\begin{eqnarray}
  \partial_{x_{\rho}} \left[ \sum_{x_{\text{op}}}  \frac{1}{2} \epsilon_{i, j,
  k} \left( x_{\text{op}} - x_{\tmop{ref}} \right)_j i\mathcal{H}^C_{\rho,
  \sigma, \kappa, k} (x, y, z, x_{\tmop{op}}) \right] & = & 0.
\end{eqnarray}
Based on arguments similar to those given in Eqs. (22)-(24) of Ref.~{\cite{Blum:2015gfa}}, we conclude that
\begin{eqnarray}
  \sum_x \left[ \sum_{x_{\text{op}}}  \frac{1}{2} \epsilon_{i, j, k} \left(
  x_{\text{op}} - x_{\tmop{ref}} \right)_j i\mathcal{H}^C_{\rho, \sigma,
  \kappa, k} (x, y, z, x_{\tmop{op}}) \right] & = & 0,
  \label{eq:muon-line-div-zero}
\end{eqnarray}
provided surface terms are neglected.
Similar conclusions hold for the sums over $y$ and $z$ as well. This
implies that
\begin{eqnarray}
  \sum_{x, y, z} \mathfrak{G}^{(1)}_{\sigma, \kappa, \rho} (y, z, z) \left[
  \sum_{x_{\text{op}}}  \frac{1}{2} \epsilon_{i, j, k} \left( x_{\text{op}} -
  x_{\tmop{ref}} \right)_j i\mathcal{H}^C_{\rho, \sigma, \kappa, k} (x, y, z,
  x_{\tmop{op}}) \right] & = & 0.  \label{eq:muon-line-sub-zero}
\end{eqnarray}
{This equation demonstrates that if we substitute the subtraction terms
defined in Eq.~(\ref{eq:muon-line-v2})
back through Eqs.~(\ref{eq:muon-line-sym}) and (\ref{eq:lbl-amp})
and finally into Eq. (\ref{eq:f2-lbl-moment}), which gives their contribution
to the anomalous moment,
we will obtain zero.}
Since we use Eq.~(\ref{eq:muon-line-sym}) to obtain the QED
weighting function, the symmetry between $x$, $y$, $z$ is not
affected by the definition of $\mathfrak{G}^{(2)}_{\sigma, \kappa, \rho}
(y, z, x)$, and Eq.~(\ref{eq:f2-lbl-moment-short-z}) can still be derived for this new function.

The neglect of surface terms in Eqs. (\ref{eq:muon-line-div-zero}) and (\ref{eq:muon-line-sub-zero})
and our use of a non-conserved, local current implies that Eq. (\ref{eq:muon-line-sub-zero})
strictly holds only in the infinite-volume and continuum limits.
In other words, for finite volume or
non-zero lattice spacing, $\mathfrak{G}^{(1)}_{\sigma, \kappa, \rho} (y, z,
x)$ and $\mathfrak{G}^{(2)}_{\sigma, \kappa, \rho} (y, z, x)$ are subject to
different finite volume and lattice spacing effects.
Lattice fermion propagators are different from their continuum counterparts
mostly in the short-distance region where the source and sink coordinates are the same,
or separated by only a few lattice spacings.
Hence the dominant discretization errors most likely come from this region.
Since the new QED weighting function satisfies Eq.~(\ref{eq:muon-line-v2-prop}),
it will suppress the contribution from this region along with its associated discretization error.
As a result, we expect smaller discretization effects if we switch
from $\mathfrak{G}^{(1)}_{\sigma, \kappa, \rho} (y, z, x)$ to
$\mathfrak{G}^{(2)}_{\sigma, \kappa, \rho} (y, z, x)$.
As we shall see in Sec. \ref{sec:results}, the new QED weighting function
indeed generates a smaller discretization error, and fortunately, a smaller finite volume
error as well.

\section{Results}\label{sec:results}

Following Ref. {\cite{Blum:2015gfa}}, we test this new framework by performing a
pure QED light-by-light calculation where the analytic result is well known~\cite{Laporta:1991zw,Laporta:1992pa,Kuhn:2003pu}.
That is, we replace the quark propagators
in Eq.~(\ref{eq:hadronic-loop}) with a leptonic loop. In Ref.
{\cite{Blum:2015gfa}}, we {studied the case where the lepton loop
mass was equal to the muon mass, $m = m_{{\mu}}$}. In this study, we also investigate the case
where the loop mass is two times the muon mass, $m = 2 m_{{\mu}}$. For both
cases, we compare results for weighting functions $\mathfrak{G}^{(1)}_{\sigma,
\kappa, \rho} (y, z, x)$ and $\mathfrak{G}^{(2)}_{\sigma, \kappa, \rho} (y, z,
x)$.

\begin{figure}[!htbp]
  \begin{center}
    \resizebox{0.4\columnwidth}{!}{\includegraphics[trim={0 0 0 1.8cm},clip]{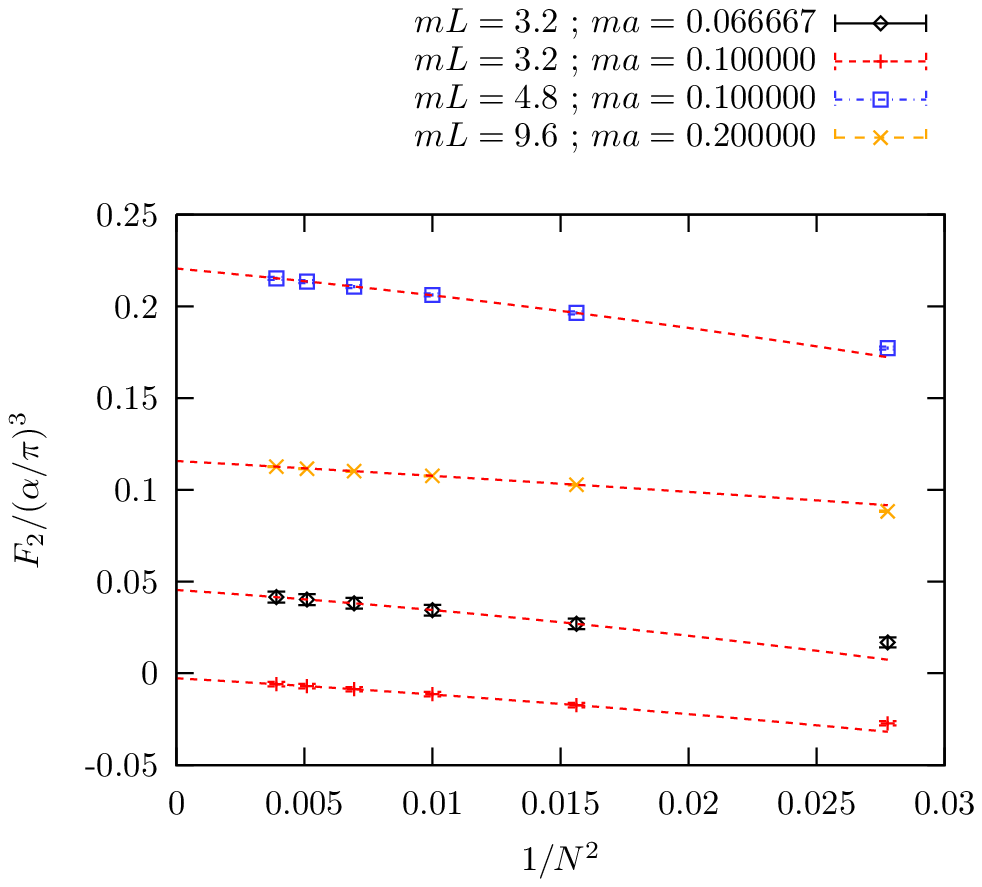}}
    \resizebox{0.4\columnwidth}{!}{\includegraphics{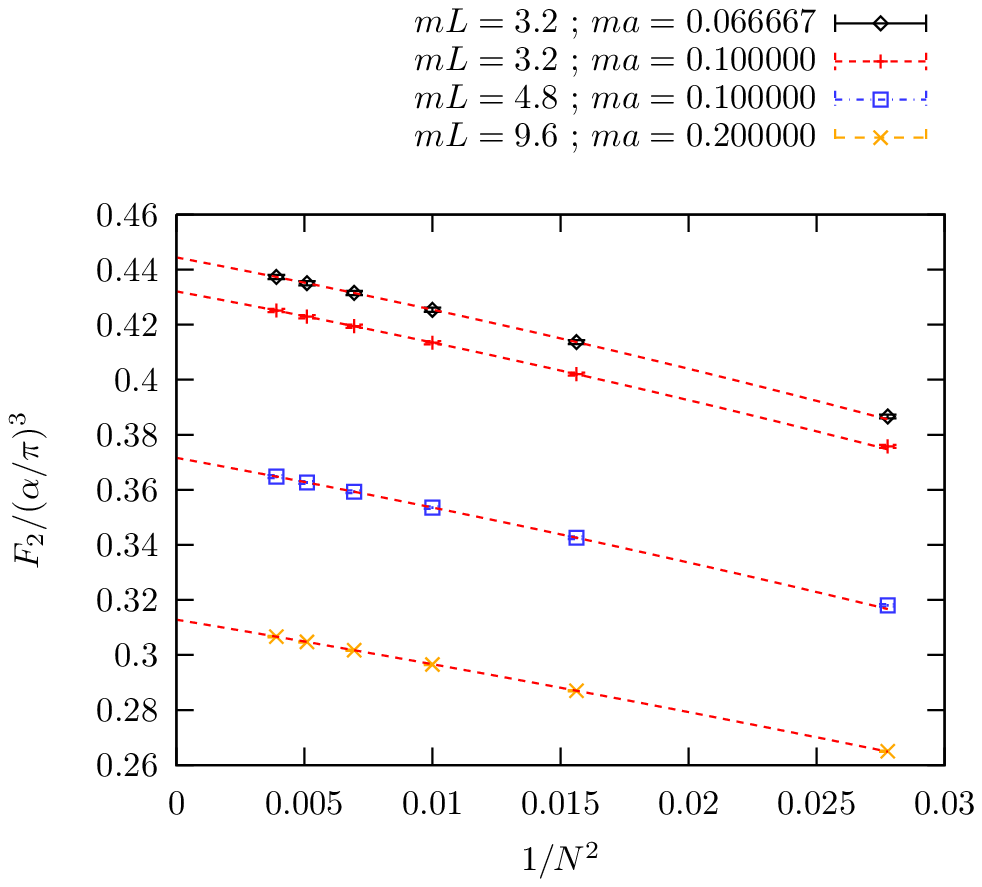}}

    \resizebox{0.4\columnwidth}{!}{\includegraphics[trim={0 0 0 1.8cm},clip]{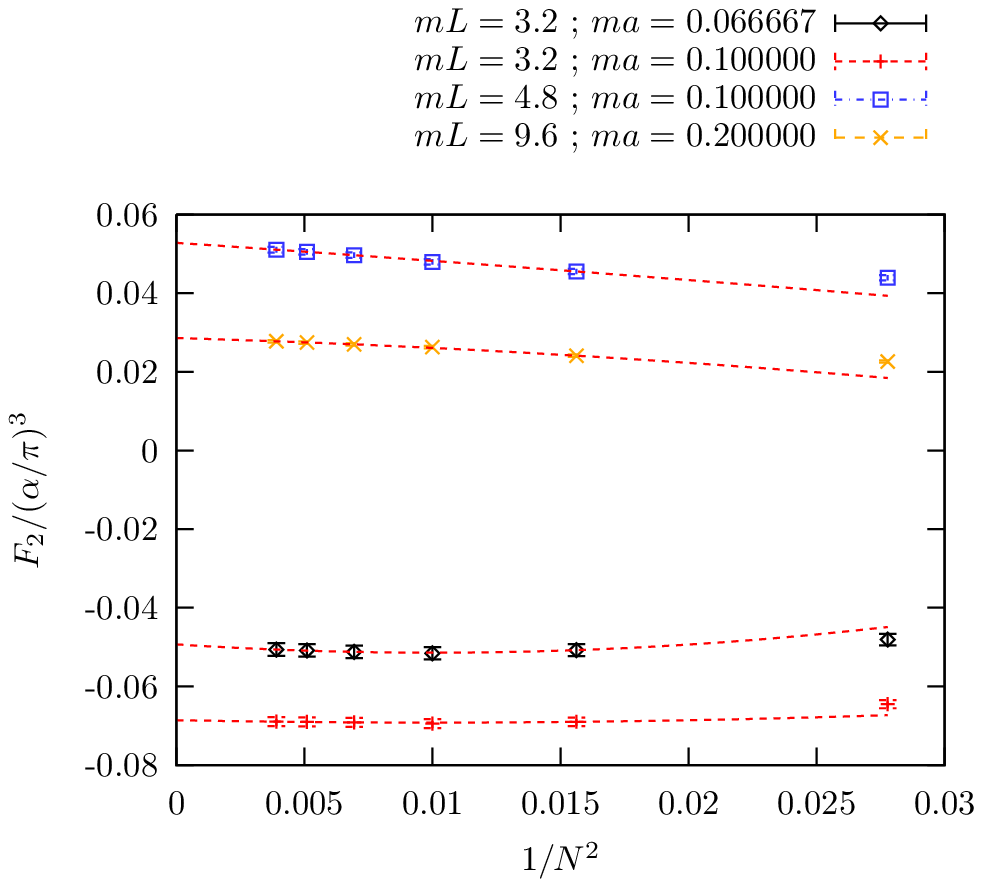}}
    \resizebox{0.4\columnwidth}{!}{\includegraphics[trim={0 0 0 1.8cm},clip]{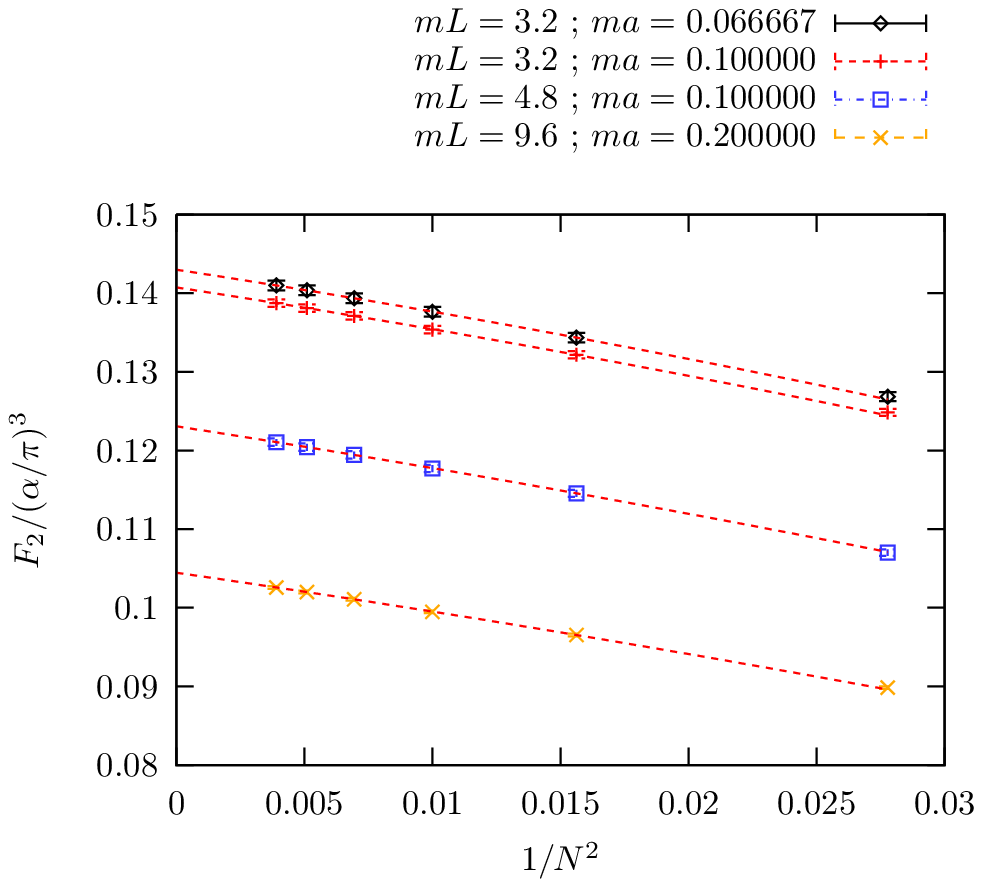}}
  \end{center}
  \caption{\label{fig:interp-fig}Extrapolations taking {the number of interpolation grid points} $N\to\infty$ for various lattices used in this study. The
  six points for each volume and lattice spacing correspond to $N = 6,
  8, 10, 12, 14, 16$. The curves are {second-order fits to $1 / N^2$}, based
  on the three points $N = 8, 12, 16$. The upper two plots
  correspond to $m = m_{{\mu}}$, the lower two, $m =
  2 m_{{\mu}}$. The left two plots correspond to $\mathfrak{G}^{(1)}$ and
  the right two plots correspond to $\mathfrak{G}^{(2)}$.}
\end{figure}

We compute $F_2$ from Eq.~(\ref{eq:f2-lbl-moment-short-z}). As mentioned
before, sums over $\tilde{x}_{\text{op}}$ and $\tilde{z}$ are performed
over the complete lattice volume, but the sum over $r$ is performed stochastically by sampling
$x$-$y$ point pairs. In order to reduce the statistical uncertainty
from this stochastic process, we sample all pairs with $r
\leqslant 6$ in lattice units, up to discrete symmetries. These amount to
183 $x$-$y$ pairs. For $r > 6$, we sample $r$ with the following
empirically chosen distribution.
\begin{eqnarray}
  p (r) & \propto & \frac{e^{- 2 m | r |}}{| r |^3} .
\end{eqnarray}
In all the cases presented below, we sampled $4096$ pairs with $r > 6$. For each
pair, we compute $F_2$ with the corresponding pre-computed, interpolated, function $i^3
\mathcal{G}_{\rho, \sigma, \kappa} (x, y, z)$ with grid sizes $N = 6,
8, 10, 12, 14, 16$. The $F_2$ values for different grids are strongly
correlated. We extrapolate to $N \rightarrow \infty$ with a second-order
fit in $1 / N^2$, using three values with $N = 8, 12, 16$. In Fig.
\ref{fig:interp-fig}, we plot fit curves corresponding to typical volumes and lattice spacings for the lepton loop.

\begin{figure}[!htbp]
  \begin{center}
    \resizebox{0.45\columnwidth}{!}{\includegraphics[trim={0 0 0 1.8cm},clip]{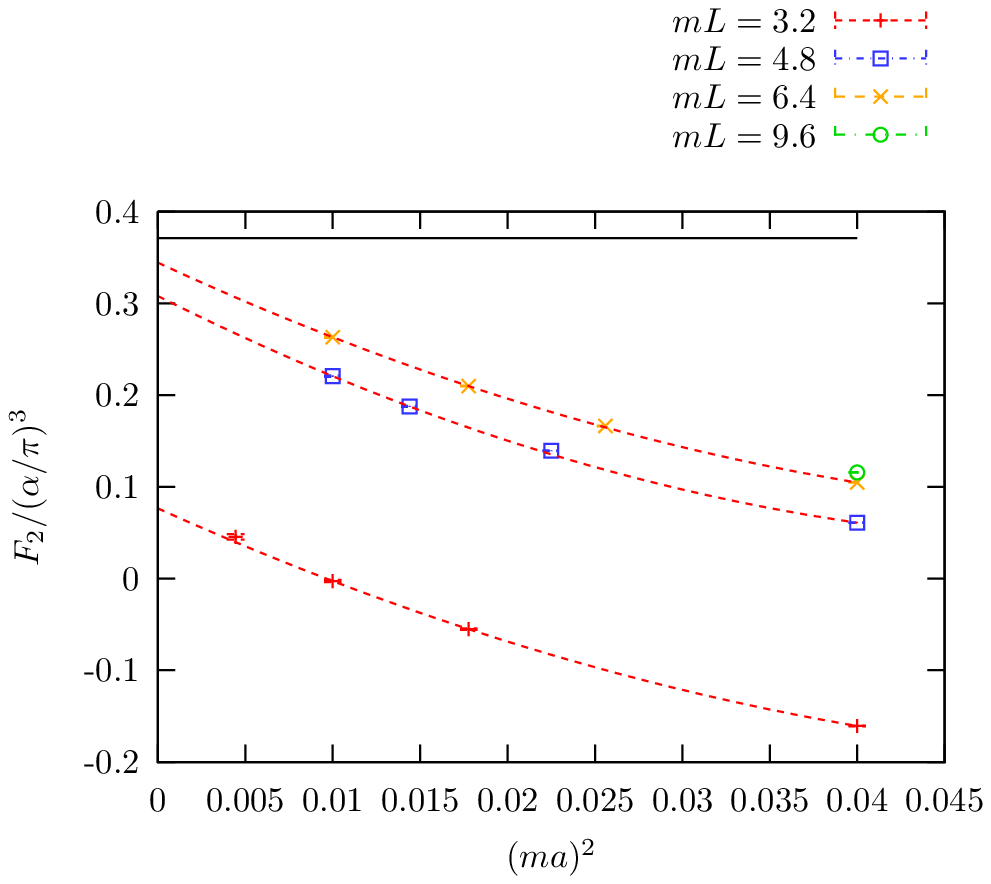}}
    \resizebox{0.45\columnwidth}{!}{\includegraphics{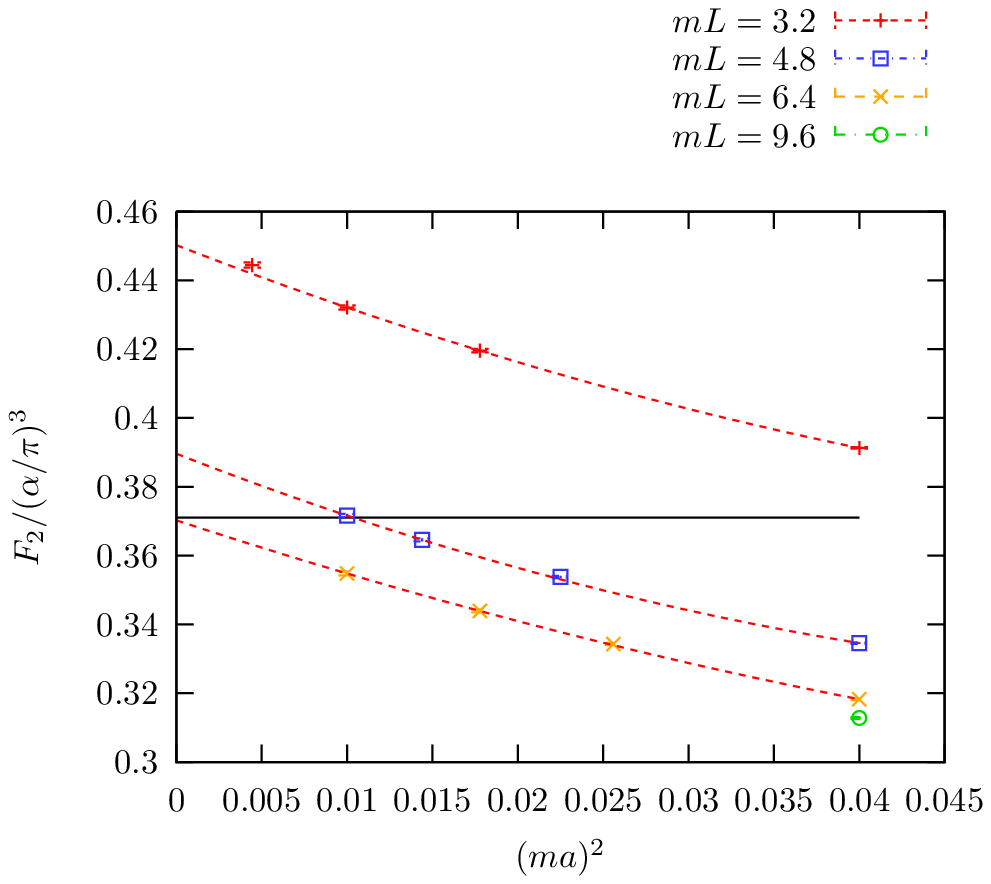}}

    \resizebox{0.45\columnwidth}{!}{\includegraphics[trim={0 0 0 1.8cm},clip]{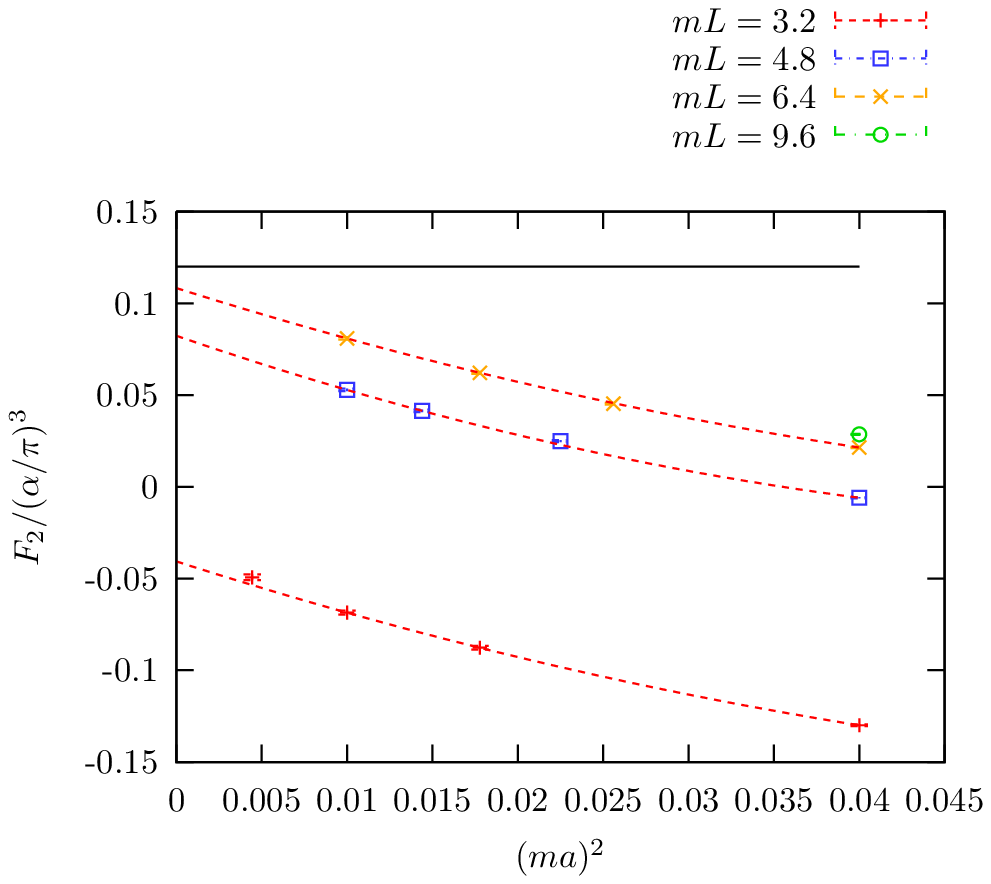}}
    \resizebox{0.45\columnwidth}{!}{\includegraphics[trim={0 0 0 1.8cm},clip]{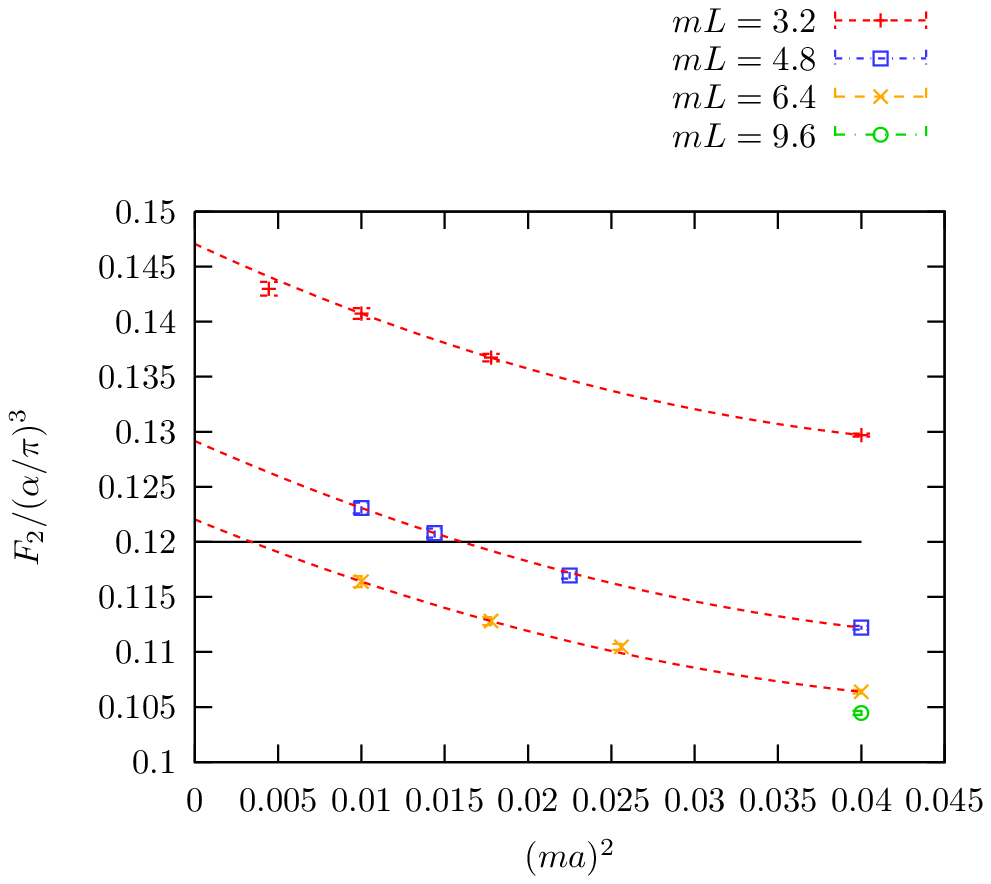}}
  \end{center}
  \caption{\label{fig:continuum-plot}Leptonic light-by-light
  contribution to the muon anomaly, with the lepton loop mass $m =
  m_{{\mu}}$ (upper) and $m = 2 m_{{\mu}}$ (lower).
  The continuum, infinite volume, result is $0.371 \times (\alpha / \pi)^3$ for $m
  = m_{{\mu}}$ \cite{Laporta:1991zw} and $0.120 \times (\alpha / \pi)^3$ for $m = 2
  m_{{\mu}}$ \cite{Laporta:1992pa,Kuhn:2003pu}. The lefthand plots correspond to
  $\mathfrak{G}^{(1)}_{\sigma, \kappa, \rho} (y, z, x)$ and the
  righthand to $\mathfrak{G}^{(2)}_{\sigma,
  \kappa, \rho} (y, z, x)$. For each volume, we draw a second-order line which
  exactly passes through the three points with $ma=0.1$, 0.12 $\tmop{or}
  0.133333$, and 0.2 to guide the eye. Note that the vertical scales between the plots on the
  left and right are different. The discretization error observed on the left is
  larger than on the right by a factor of four, or more, while the finite volume errors are larger by a factor of two, or more.
  The parameters for these curves are given in Tab. \ref{tab:plot-lines}.}
\end{figure}

After removing the interpolation error for $i^3 \mathcal{G}_{\rho, \sigma,
\kappa} (x, y, z)$, we study non-zero lattice spacing and
finite volume effects. The results are plotted in Fig. \ref{fig:continuum-plot},
and the parameter values are listed in Tab. \ref{tab:plot-lines}.
The finite volume and non-zero lattice spacing effects are much reduced by
using $\mathfrak{G}^{(2)}$ instead of $\mathfrak{G}^{(1)}$, and
the curves for different volumes appear to be quite parallel. Note that in the latter case some points even have the wrong sign.
The difference between
the $m a = 0.1$ and $ma = 0.2$ results is a good indicator of the non-zero
lattice spacing effects. Since we have obtained results for $ma = 0.1$ and $0.2$ for three
volumes, this difference demonstrates the volume dependence
of the non-zero lattice spacing effects. We show this comparison in Tab.
\ref{tab:vol-spacing}.
The $mL = 4.8$ and $6.4$ points agree within errors for both loop masses. The volume $mL = 3.2$ shows similar effects, but in some cases, given our high statistical precision, we observe a small difference.
{
This is expected since the non-zero lattice spacing effects become independent of volume in the large volume limit.
We also study the lattice spacing dependence of the finite volume effects in Tab. \ref{tab:spacing-vol}.
It can be seen from the table that the finite volume effects are roughly independent of lattice spacing.
The finite volume effects at fixed lattice spacing $ma = 0.2$ are shown in Tab. \ref{tab:vol-0.2},
and we expect that the finite volume effects in the continuum limit are similar.
With this table, we can see that the finite volume effect, falling exponentially
with the linear size of the lattice,
becomes negligible for $mL = 9.6$.
}

\begin{table}[h]
  \caption{\label{tab:plot-lines}Fits {of the muon anomaly after taking the number of interpolation grid points $N\to\infty$ }for non-zero lattice spacing shown Fig. \ref{fig:continuum-plot}.}
  \begin{center}
    \begin{tabular}{cccc}
      \hline
      \hline
      $m / m_{{\mu}}$ & $mL$ & $F_2 / (\alpha / \pi)^3$ using
      $\mathfrak{G}^{(1)}$ & $F_2 / (\alpha / \pi)^3$ using
      $\mathfrak{G}^{(2)}$\\
      \hline
      1 & 3.2 & $0.0765 (41) - 8.58 (37) (ma)^2 + 66 (7) (ma)^4$ & $0.4502 (23)
      - 1.92 (22) (ma)^2 + 11.3 (4.1) (ma)^4$\\
      1 & 4.8 & $0.3080 (43) - 9.59 (44) (ma)^2 + 85 (9) (ma)^4$ & $0.3896 (27)
      - 1.94 (28) (ma)^2 + 14.1 (5.3) (ma)^4$\\
      1 & 6.4 & $0.3443 (26) - 8.83 (23) (ma)^2 + 71 (5) (ma)^4$ & $0.3703 (19)
      - 1.63 (18) (m a)^2 + 8.2 (3.4) (m a)^4$\\
      \hline
      2 & 3.2 & $- 0.0407 (41) - 2.98 (40) (ma)^2 + 19 (8) (ma)^4$ & $0.1471
      (18) - 0.70 (16) (ma)^2 + 6.6 (3.0) (ma)^4$\\
      2 & 4.8 & $0.0823 (39) - 3.20 (42) (ma)^2 + 25 (8) (ma)^4$ & $0.1292 (26)
      - 0.67 (27) (ma)^2 + 6.2 (5.2) (ma)^4$\\
      2 & 6.4 & $0.1083 (23) - 2.94 (22) (ma)^2 + 19 (5) (ma)^4$ & $0.1220 (17)
      - 0.62 (16) (m a)^2 + 5.8 (2.9) (m a)^4$\\
      \hline
      \hline
    \end{tabular}
  \end{center}
\end{table}

\begin{table}[h]
  \caption{\label{tab:vol-spacing}{Volume dependence of non-zero lattice
  spacing effects in the muon anomaly. Differences between $F_2$ at $ma = 0.1$ and $ma = 0.2$ are shown
  for each volume.}}
  \begin{center}
    \begin{tabular}{cccc}
      \hline
      \hline
      $m / m_{{\mu}}$ & $mL$
      & $\Delta F_2 / (\alpha / \pi)^3$ using $\mathfrak{G}^{(1)}$
      & $\Delta F_2 / (\alpha / \pi)^3$ using $\mathfrak{G}^{(2)}$\\
      \hline
      1 & 3.2 & $0.1580 (13)$ & $0.0408 (7)$\\
      1 & 4.8 & $0.1597 (9)$ & $0.0370 (6)$\\
      1 & 6.4 & $0.1584 (8)$ & $0.0365 (6)$\\
      \hline
      2 & 3.2 & $0.0614 (13)$ & $0.0110 (5)$\\
      2 & 4.8 & $0.0588 (8)$ & $0.0109 (5)$\\
      2 & 6.4 & $0.0594 (8)$ & $0.0100 (6)$\\
      \hline
      \hline
    \end{tabular}
  \end{center}
\end{table}

\begin{table}[h]
  \caption{\label{tab:spacing-vol}{Lattice spacing dependence of finite volume effects for the muon anomaly.
  Differences between $F_2$ with lattice size $L$ and $6.4/m$ are shown
  for two lattice spacings, $ma = 0.1$ and $ma = 0.2$.}}
  \begin{center}
    \begin{tabular}{ccccccc}
      \hline
      \hline
      $m / m_{{\mu}}$& $m L$ & $m a$ 
      & $\Delta F_2 / (\alpha / \pi)^3$ using $\mathfrak{G}^{(1)}$
      & $\Delta F_2 / (\alpha / \pi)^3$ using $\mathfrak{G}^{(2)}$\\
      \hline
      1 & 3.2 & 0.1 & $-0.2658(15)$ & $0.0773 (9)$\\
      1 & 3.2 & 0.2 & $-0.2654 (5)$ & $0.0730 (4)$\\
      1 & 4.8 & 0.1 & $-0.0425(12)$ & $0.0168 (8)$\\
      1 & 4.8 & 0.2 & $-0.0438 (4)$ & $0.0163 (4)$\\
      \hline
      2 & 3.2 & 0.1 & $-0.1494(14)$ & $0.0243 (7)$\\
      2 & 3.2 & 0.2 & $-0.1514 (6)$ & $0.0233 (3)$\\
      2 & 4.8 & 0.1 & $-0.0280(10)$ & $0.0067 (7)$\\
      2 & 4.8 & 0.2 & $-0.0274 (5)$ & $0.0058 (3)$\\
      \hline
      \hline
    \end{tabular}
  \end{center}
\end{table}

\begin{table}[h]
  \caption{\label{tab:vol-0.2}Volume dependence of the muon anomaly at fixed non-zero lattice spacing. $ma = 0.2$. It
  can be seen that the infinite volume value can be approximated by the
  largest volume ($mL = 9.6$) result. The column ``diff'' is the finite volume
  effect at this volume and lattice spacing, calculated by taking the
  difference between $F_2 / (\alpha / \pi)^3$ given in that row and in the $mL
  = 9.6$ row.}
  \begin{center}
    \begin{tabular}{cccccc}
      \hline
      \hline
      $m / m_{{\mu}}$ & $mL$ & $F_2 / (\alpha / \pi)^3$ using
      $\mathfrak{G}^{(1)}$ & diff & $F_2 / (\alpha / \pi)^3$ using
      $\mathfrak{G}^{(2)}$ & diff\\
      \hline
      1 & 3.2 & $- 0.1607 (4)$ & $- 0.2765 (5)$ & $0.3913 (3)$ & $0.0785 (4)$\\
      1 & 4.8 & $0.0609 (3)$ & $- 0.0548 (4)$ & $0.3346 (3)$ & $0.0217 (4)$\\
      1 & 6.4 & $0.1047 (3)$ & $- 0.0110 (4)$ & $0.3182 (3)$ & $0.0054 (4)$\\
      1 & 9.6 & $0.1157 (4)$ & 0 & $0.3128 (3)$ & 0\\
      \hline
      2 & 3.2 & $- 0.1300 (5)$ & $- 0.1586 (6)$ & $0.1297 (2)$ & $0.0252 (3)$\\
      2 & 4.8 & $- 0.0060 (4)$ & $- 0.0346 (5)$ & $0.1122 (2)$ & $0.0077 (3)$\\
      2 & 6.4 & $0.0214 (4)$ & $- 0.0072 (5)$ & $0.1064 (2)$ & $0.0019 (3)$\\
      2 & 9.6 & $0.0286 (4)$ & 0 & $0.1044 (2)$ & 0\\
      \hline
      \hline
    \end{tabular}
  \end{center}
\end{table}

Since the finite volume effects are exponentially suppressed with lattice size $L$ and the non-zero lattice spacing effects are of order
$a^2$, the lepton anomaly scales like
\begin{eqnarray}
  F_2 (L, a) & = & F_2 +\mathcal{O} (e^{- mL}) +\mathcal{O} ((ma)^2) .
\end{eqnarray}
So far, from Tab. \ref{tab:vol-spacing} and Tab. \ref{tab:vol-0.2}, we have made
two observations: 1) the non-zero lattice spacing effect becomes {approximately}
independent of volume when $mL \geqslant 4.8$; 2) the finite volume effect
becomes negligible for $mL = 9.6$. Based on these two observations, we
fit all of the $mL \geqslant 4.8$ data with the following second-order
formula:
\begin{eqnarray}
  F_2 (L, a) & = & F_2 (L) + k_1 a^2 + k_2 a^4 .
\end{eqnarray}
To study the systematic effects, we also fit the data with a third-order
formula:
\begin{eqnarray}
  F_2 (L, a) & = & F_2 (L) + k_1 a^2 + k_2 a^4 + k_3 a^6 .
\end{eqnarray}
We do not assume any specific functional form of $F_2 (L)$. Instead, we assume
\begin{eqnarray}
  F_2 & \approx & F_2 (9.6 / m) .
\end{eqnarray}

In this scheme, we show final results for two fermion loop masses and for
$\mathfrak{G}^{(1)}$ and $\mathfrak{G}^{(2)}$ in Tab. \ref{tab:final-fit}. We
can see that our method, with the third-order fit, has successfully reproduced the analytic calculation within our statistical precision
in all cases. For $\mathfrak{G}^{(1)}$ the 2nd order fits disagree outside of statistical errors, but the values are still quite
close, within five percent or less. Using 3rd order fits and $\mathfrak{G}^{(2)}$ for central values, and the difference between 2nd and 3rd order fits as a systematic error, we find
\begin{eqnarray}
F_2 / (\alpha / \pi)^3 &=& 0.3686(37)(35), \\
F_2 / (\alpha / \pi)^3 &=& 0.1232(30)(28),
\end{eqnarray}
for $m/m_\mu=1$ and $2$, respectively. {Here, the first error is statistical and the second systematic.}
These values agree within one standard deviation to the analytic results~\cite{Laporta:1991zw,Laporta:1992pa,Kuhn:2003pu}, 0.371 and 0.120, for the two loop masses.

\begin{table}[h]
  \caption{\label{tab:final-fit}The muon anomaly in the continuum and infinite volume from fits to values with $mL =
  4.8, 6.4, 9.6$. Results are given for 2nd order ($F_2 (L) + k_1 a^2 + k_2 a^4$) and
  3rd order ($F_2 (L) + k_1 a^2 + k_2 a^4 + k_3 a^6$) fits. ``dof'' denotes degrees of freedom, and $\chi^2$ is an uncorrelated chi-squared value from the fit. The analytic results are computed using continuum, infinite volume, perturbation theory~\cite{Laporta:1991zw,Laporta:1992pa,Kuhn:2003pu}.}
  \begin{center}
    \begin{tabular}{ccccccc}
      \hline
      \hline
      $m / m_{{\mu}}$ & order & dof & $F_2 / (\alpha / \pi)^3$ using
      $\mathfrak{G}^{(1)}$ & $\chi^2$ & $F_2 / (\alpha / \pi)^3$ using
      $\mathfrak{G}^{(2)}$ & $\chi^2$\\
      \hline
      1 & 2 & $9 - 5 = 4$ & $0.3522 (14)$ & 11.3 & $0.3651 (10)$ & 2.5\\
      1 & 3 & $9 - 6 = 3$ & $0.3647 (51)$ & 2.8 & $0.3686 (37)$ & 1.4\\
      \hline
      1 & analytic &  &  &  & & $0.371$  \\
      \hline
      2 & 2 & $9 - 5 = 4$ & $0.1146 (13)$ & 3.6 & $0.1204 (9)$ & $4.5$\\
      2 & 3 & $9 - 6 = 3$ & $0.1153 (44)$ & 3.6 & $0.1232 (30)$ & $3.6$\\
      \hline
      2 & analytic &  &  &  & &$0.120$  \\
      \hline
      \hline
    \end{tabular}
  \end{center}
\end{table}

\begin{figure}[!htbp]
  \begin{center}
    \resizebox{0.7\columnwidth}{!}{\includegraphics{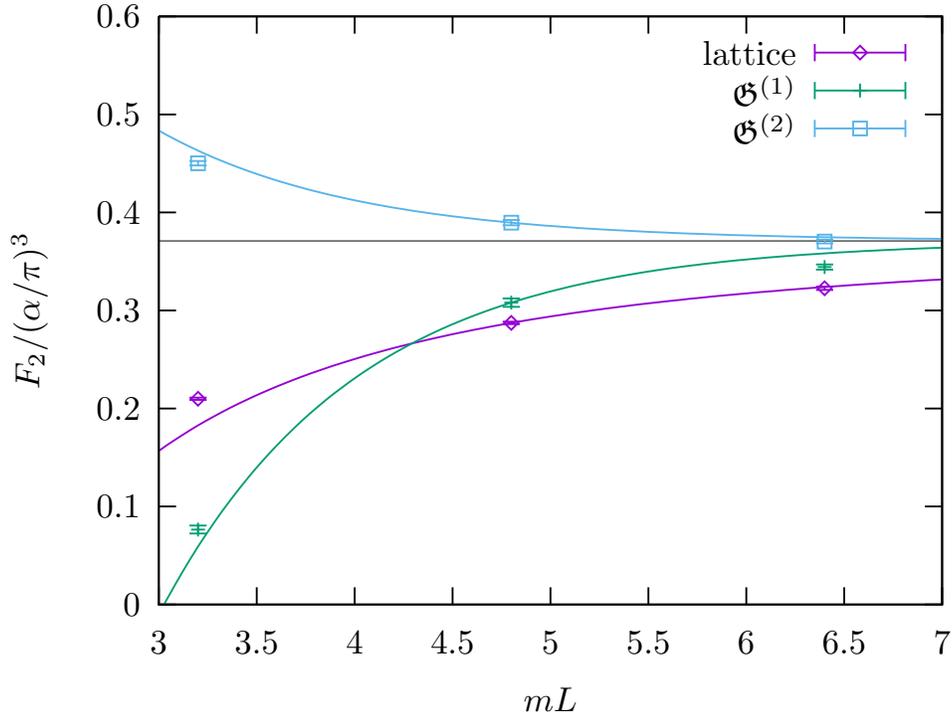}}
  \end{center}
  \caption{\label{fig:finite-vol-dep}Volume dependence of the muon anomaly for infinite and finite volume QED weighting functions.
  The diamonds correspond to
  the finite volume QED weighting function computed on the lattice~\cite{Blum:2015gfa}.
  The plus signs and squares correspond to
  infinite volume QED weighting functions $\mathfrak{G}^{(1)}$ and $\mathfrak{G}^{(2)}$, respectively.
  Values are listed in Tab.~\ref{tab:plot-lines}.
  Curves correspond to expected finite volume scaling ($0.371 + k / L^2$) and
  infinite volume scaling ($0.371 + k e^{-m L}$), where the coefficient $k$ is chosen to match the data at $mL=4.8$.
  The right most point for the finite volume weighting function lies a bit off its scaling curve
   because the discretization error has not been completely removed by the
   simple ansatz given in Tab.~\ref{tab:plot-lines}, and the coefficient $k$ does not contain any possible volume dependence.
}
\end{figure}

Finally, to illustrate how exponentially-suppressed finite volume errors compare with the
power-law suppressed finite volume effects seen in Ref. \cite{Blum:2015gfa},
we show the values from Tab.~XI in Ref. \cite{Blum:2015gfa}
and from Tab. \ref{tab:plot-lines} in Fig. \ref{fig:finite-vol-dep}.
The curves shown in the figure, which are not fits, demonstrate the expected volume dependence of
the old finite volume QED weighting function and the new infinite volume one.
The simple scaling curves also do not account for possible volume dependence of pre-factors.
The {rightmost green, plus sign} point for the infinite-volume weighting function
$\mathfrak{G}^{(1)}$
lies a bit off the corresponding curve.
This most likely results because the discretization error has not been completely removed by the
simple ansatz given in Tab.~\ref{tab:plot-lines}.
This is confirmed in Tab.~\ref{tab:final-fit},
where for $m/m_\mu = 1$ the 2nd and 3rd order fit values for $\mathfrak{G}^{(1)}$ do not agree well.
Note the 2nd order fit is especially poor.
Still, we can clearly see that the curves for the infinite volume QED weighting functions
approach the analytic result much faster than the curve for the finite volume QED weighting function, as expected.

\section{Conclusion}

In this paper we outlined an approach to eliminate the $1 / L^2$ finite
volume errors in previous hadronic light-by-light calculations
{\cite{Blum:2016lnc,Blum:2015gfa}}. This work was very much motivated by the recent progress made in Ref. {\cite{Asmussen:2016lse}}.
In comparison, our approach requires less analytic calculation but more
numerical effort. Since we do not average over the direction of the propagating muon line,
our function $i^3 \mathcal{G}_{\rho, \sigma, \kappa} (x, y, z)$
depends on five parameters instead of three, which makes the interpolation harder.
However, as we have demonstrated in Sec.
\ref{sec:results}, these difficulties have been overcome. We noticed that one has
 freedom in choosing the QED weighting function $i^3 \mathcal{G}_{\rho,
\sigma, \kappa} (x, y, z)$ without affecting the final result. This added freedom can
potentially reduce the discretization and finite volume errors.
In particular, we find that the choice $\mathfrak{G}^{(2)}$
defined by Eq.~(\ref{eq:muon-line-v2}) is much better than the original
$\mathfrak{G}^{(1)}$ defined in Eqs.~(\ref{eq:muon-line-v1}) and (\ref{eq:muon-line}).
We are now applying the new infinite volume QED weighting function $i^3
\mathcal{G}_{\rho, \sigma, \kappa} (x, y, z)$ obtained in this work to the
hadronic four-point correlation function already computed (and saved) in our previous work
{\cite{Blum:2016lnc}}.

\section{Acknowledgements}

We thank our RBC and UKQCD collaborators for helpful discussions
and support.
The BAGEL~\cite{Boyle:2009vp} library was used to verify the code
to compute free DWF propagators with fast Fourier transforms
using FFTW~\cite{Frigo:1999:FFT:301618.301661}.
The CPS~\cite{Jung:2014ata}\footnote{The CPS software repository can be found at \texttt{https://github.com/RBC-UKQCD/cps}.} software package is also used in the calculation.
In addition, we thank RBRC for BG/Q computer time.
Computations were mainly performed under the ALCC Program of the US DOE on the
Blue Gene/Q (BG/Q) Mira computer at the Argonne Leadership Class Facility, a
DOE Office of Science Facility supported under Contract De-AC02-06CH11357.
Computations were also supported through resources provided by the Scientific Data and Computing Center (SDCC) at Brookhaven National Laboratory (BNL), a DOE Office of Science User Facility supported by the Office of Science of the U.S. Department of Energy. The SDCC is a major component of the Computational Science Initiative (CSI) at BNL.
L.C.J. is supported by the Department of Energy, Laboratory Directed Research and Development (LDRD) funding of BNL, under contract DE-SC0012704.
T.I., C.J., and C.L. are supported in part by US DOE Contract DE-SC0012704(BNL).
T.I. is supported in part by JSPS KAKENHI Grant Numbers JP26400261, JP17H02906, and also is supported by MEXT as "Priority Issue on Post-K computer" (Elucidation of the Fundamental Laws and Evolution of the Universe) and JICFuS.
T.B. is supported by US DOE grant DE-FG02-92ER40716.
N.H.C. is supported in part by US DOE grant \#DE-SC0011941.
M.H. is supported in part by Japan Grants-in-Aid for Scientific Research, No.16K05317.
C.L. acknowledges support through a DOE Office of Science Early Career Award.

\bibliographystyle{unsrt}
\bibliography{ref.bib}

\end{document}